\documentclass[twocolumn,showpacs,preprintnumbers,amsmath,amssymb,fleqn]{revtex4}
\usepackage{graphics}
\normalsize

\begin{document}
\title{Controlling entanglement sudden death in cavity QED by classical driving fields}
\author{Jian-Song Zhang} \author{Jing-Bo Xu}%
 \email{xujb@zju.edu.cn}
\affiliation{Zhejiang Institute of Modern Physics and Physics Department,\\
Zhejiang University, Hangzhou 310027, People's Republic of China }

\author{Qiang Lin}
\affiliation{Institute of Optics, Department of Physics, Zhejiang
University, Hangzhou 310027, China}

\date{\today}

\begin{abstract}
 We investigate the entanglement dynamics of a quantum system consisting of two-level
 atoms interacting with vacuum or thermal fields with classical driving fields.
 We find that the entanglement of
the system can be improved  by adjusting the classical driving
field. The influence of the classical field and the purity of the
initial state on the entanglement sudden death is also studied. It
is shown that the time of entanglement sudden death can be
controlled by the classical driving fields. Particularly, the
entanglement sudden death phenomenon will disappear if the classical
driving fields are strong enough.
\end{abstract}
\pacs{03.67.Mn; 03.65.Ud}
 \maketitle

\section{INTRODUCTION}
Entanglement, one of the most striking features of quantum
mechanics, has been considered as a key resource of quantum
information processing
\cite{Cirac2000,Nielsen2000,Bennett1993,vedral2000}. In recent
years, the manipulation of quantum entanglement for the system of
the cavity quantum electrodynamics(QED) has been extensively
investigated\cite{Raimond2001,Bose2001,Kim2002,Paternostro2007,Vitali2007,
Li2005,Marchiolli2006,Ke2005,Braun2002,Xu2005}. The cavity QED,
where atoms interacting with quantized electromagnetic fields inside
a cavity, can be used to create the entanglement between atoms in
cavities and establish quantum communications between different
optical cavities. It has been shown that the entanglement can arise
in the interaction of a atom with a cavity field in a thermal state
\cite{Bose2001, Kim2002} and two atoms with a cavity field
\cite{Braun2002, Xu2005}. In Ref.\cite{Solano2003}, Solano \emph{et
al}. have shown that multipartite entanglement can be generated by
putting several two-level atoms in a cavity of high quality factor.
With the help of a strong classical driving fields, the
Schr\"{o}dinger cat state and other entangled states can be
produced.

On the other hand, many efforts have been devoted to the study of
the evolution of the joint system formed by two qubits\cite{Yu2004,
Jakobczyk2004, Ficek2006, Yonac, Yu2006, Ikram2008, Bellomo2007,
Bellomo2008}. Particularly, Yu and Eberly \cite{Yu2004} pointed out
that the single-qubit dynamics and the global dynamics of an
entangled two-qubit system subjected to independent environments may
be rather different. For example, for a single-qubit system
subjected to an environment, the local coherence decays
asymptotically. However, the entanglement of an entangled two-qubit
system may disappear within a finite time during the dynamics
evolution. The nonsmooth finite-time disappearance of entanglement
is called ``entanglement sudden death'' (ESD). Recently, it is
reported that the ESD phenomenon has been observed in a quantum
optics experiment \cite{Almeida2007}.

In the present paper, we propose a scheme to enhance the
entanglement of a quantum system consisting of two-level atoms
within cavities by applying and controlling classical driving
fields. The influence of the classical field and the purity of the
initial state on the ESD is also studied. First, we study the
entanglement of the two-level atom and the field by employing
logarithmic negativity \cite{Vidal2002, Plenio2005}. Then, we
consider a quantum system consisting of two noninteracting atoms
each locally interacting with its own vacuum field. The two atoms,
which are initially prepared in extended Werner-like (EWL) states
\cite{Bellomo2008}, are driven by two independent classical fields
additionally. It is shown that the ESD may appear in this system and
the ESD time can be retarded by increasing the purity of the initial
state of the two atoms. In addition, the amount of the entanglement
of the two atoms can be significantly increased by applying
classical fields. We find that the time of entanglement sudden death
can be controlled by the classical driving fields. It is interesting
to point out that the entanglement sudden death phenomenon disappear
if the classical driving fields are strong enough.

The present paper is organized as follows. In section II, we propose
a scheme to improve the entanglement of a two-level atom interacting
with vacuum or thermal field by applying a classical driving field.
In section III, we study the influence of the classical driving
fields on the ESD when the atoms are initially prepared in EWL
states. A conclusion is given in section IV.

\section{Entanglement dynamics of a two-level atom in a cavity with a classical driving field}
\subsection{The model}
Now, we consider a system consisting of a two-level atom inside a
single mode cavity. The atom is driven by a classical field
additionally. The Hamiltonian of the system can be described by
\cite{Solano2003}
\begin{eqnarray}
H&=&\omega
a^{\dag}a+\frac{\omega_0}{2}\sigma_z+g(\sigma_+a+\sigma_-a^{\dag})\nonumber\\
&&+\lambda(e^{-i\omega_ct}\sigma_++e^{i\omega_ct}\sigma_-),
\end{eqnarray}
where $\omega$, $\omega_0$ and $\omega_c$ are the frequency of the
cavity, atom and classical field, respectively. The operators
$\sigma_z$ and
 $\sigma_{\pm}$ are defined by  $\sigma_{z}=|e\rangle\langle e|-|g\rangle\langle
 g|$, $\sigma_+=|e\rangle\langle g|$, and $\sigma_-=\sigma_+^{\dag}$ where
$|e\rangle$ and $|g\rangle$ are the excited and ground states of the
atom. Here, $a$ and $a^{\dag}$ are the annihilation and creation
operators of the cavity; g and $\lambda$ are the coupling constants
of the interactions of the atom with the cavity and with the
classical driving field, respectively. Note that we have set
$\hbar=1$ throughout this paper.

In the rotating reference frame the Hamiltonian of the system is
transformed to the Hamiltonian $H_1$ under a unitary transformation
$U_1= \exp{(-i\omega_c t\sigma_z/2)}$
\begin{eqnarray}
H_1&=&U_1^{\dag}HU_1-iU_1^{\dag}\frac{\partial U_1}{\partial
t}\nonumber\\
&=&H_1^{(1)}+H_1^{(2)},
\end{eqnarray}
with
\begin{eqnarray}
H_1^{(1)}&=&\omega
a^{\dag}a+g(e^{i\omega_ct}\sigma_+a+e^{-i\omega_ct}\sigma_-a^{\dag}),\nonumber\\
H_1^{(2)}&=&\frac{\Delta_1}{2}\sigma_z+\lambda(\sigma_++\sigma_-),
\end{eqnarray}
and $\Delta_1=\omega_0-\omega_c$. Using the method similar to that
used in Ref.\cite{Liu2006}, diagonalizing the Hamiltonian
$H_1^{(2)}$, and neglecting the terms which do not conserve energies
(rotating wave approximation), we can recast the Hamiltonian $H_1$
as follows

\begin{eqnarray}
H_1&=&\omega
a^{\dag}a+\frac{\Omega_1\sin{\theta}}{2}(\sigma_++\sigma_-)
+g\cos^2{\frac{\theta}{2}}[e^{i\omega_ct}\nonumber\\
&&\times(-\frac{\sin{\theta}}{2}\sigma_z+\cos^2{\frac{\theta}{2}}\sigma_+-\sin^2{\frac{\theta}{2}}\sigma_-)a
+h.c],
\end{eqnarray}
with $\theta=\arctan{(\frac{2\lambda}{\Delta_1})}$. Here $h.c$
stands for Hermitian conjugation. The Hamiltonian (4) can be
diagonalized by a final unitary transformation $U_2$ with
$U_2=\exp{[\frac{i\omega_ct}{2}(\sigma_++\sigma_-)]}$. Then, we can
rewrite the Hamiltonian of the system
\begin{eqnarray}
H_2&=&\omega
a^{\dag}a+\frac{\omega'\sin{\theta}}{2}(\sigma_++\sigma_-)+g'[(-\frac{\sin{\theta}}{2}\sigma_z\nonumber\\
&&+\cos^2{\frac{\theta}{2}}\sigma_+
-\sin^2{\frac{\theta}{2}}\sigma_-)a +h.c],
\end{eqnarray}
where $\omega'=\sqrt{\Delta_1^2+4\lambda^2}+\omega_c$ and $g'=g
\cos^2{\frac{\theta}{2}}$. It is worth noting that the unitary
transformations $U_1$ and $U_2$ are both local unitary
transformations. As we known the entanglement of a quantum system
does not change under local unitary transformations
\cite{Vedral1997}. Thus, the entanglement of the system considered
here will not be changed by applying the local unitary
transformations $U_1$ and $U_2$.

\subsection{Entanglement dynamics}
We study the entanglement of the system by employing the logarithmic
negativity. For a bipartite system described by density matrix
$\rho$, the logarithmic negativity is
\begin{equation}
E(\rho)\equiv \log_2{(1+2N)}= \log_2||\rho^{T}||,
\end{equation}
where $\rho^{T}$ is the partial transpose of $\rho$ , $||\rho^{T}||$
stands for the trace norm of $\rho^{T}$ and N is negativity which is
defined by \cite{Vidal2002}
\begin{equation}
N\equiv\frac{||\rho^{T}||-1}{2},
\end{equation}
which is just the absolute value of the sum of the negative
eigenvalues of $\rho^{T}$. The logarithmic negativity has been
proven to be an operational good entanglement
measure\cite{Plenio2005}. We first assume that the atom is initially
prepared in the state
$|+\rangle=\cos{\frac{\theta}{2}}|e\rangle+\sin{\frac{\theta}{2}}|g\rangle$
and the field is in the Fock state $|n\rangle$. Then, we can find
that the state vector at time t is
\begin{eqnarray}
|\psi{(t)}\rangle&=&\alpha_n(t)|+, n\rangle+\beta_{n+1}(t)|-,
n+1\rangle,\nonumber\\
\alpha_n(t)&=&e^{i\Delta_2t/2}[\cos{(\Omega_n
t)}-\frac{i\Delta_2}{2\Omega_n}\sin{(\Omega_nt)}], \nonumber\\
\beta_n(t)&=&-ig\cos^2{\frac{\theta}{2}}\sqrt{n}e^{-i\Delta_2t/2}\sin{(\Omega_{n-1}t)}/\Omega_{n-1},\nonumber\\
\Delta_2&=&\sqrt{(\omega_0-\omega_c)^2+4\lambda^2}+\omega_c-\omega, \nonumber\\
\Omega_n&=&\sqrt{\frac{\Delta_2^2}{4}+(n+1)(g\cos^2{\frac{\theta}{2}})^2},\nonumber\\
|+\rangle&=&\cos{\frac{\theta}{2}}|e\rangle+\sin{\frac{\theta}{2}}|g\rangle,\nonumber\\
|-\rangle&=&-\sin{\frac{\theta}{2}}|e\rangle+\cos{\frac{\theta}{2}}|g\rangle,
\end{eqnarray}
Inserting the state vector into Eq.(7) leads to the logarithmic
negativity of the state $|\psi(t)\rangle$
\begin{eqnarray}
E(|\psi\rangle)=\log_2[1+2|\alpha_n(t)\beta_{n+1}(t)|].
\end{eqnarray}

Next, we consider the entanglement dynamics of the system when the
atom and the field are initially prepared in the state $|+\rangle$
and the thermal state, respectively. The initial density matrix of
the system is
\begin{eqnarray}
\rho(0)&=&|+\rangle\langle
+|\otimes(\sum_{n=0}^{\infty}p_n|n\rangle\langle
n|),\nonumber\\
p_n&=&\frac{\overline{m}^n}{(1+\overline{m})^{n+1}},
\end{eqnarray}
where $\overline{m}=1/(e^{\beta \omega}-1)$ is the mean photon
number at the inverse temperature $\beta$ and $\omega$ is the
frequency of the optical mode. A straightforward calculation shows
that the density matrix at time t is
\begin{eqnarray}
\rho(t)&=&\sum_{n=0}^{\infty}\{p_{n-1}|\beta_n(t)|^2|-\rangle\langle-|\otimes|n\rangle\langle
n|+p_n|\alpha_n(t)|^2\nonumber\\
&&\times|+\rangle\langle+|\otimes|n\rangle\langle
n|+[p_n\alpha_n(t)\beta^*_{n+1}(t)\nonumber\\
&&\times|+\rangle\langle-|\otimes|n\rangle\langle
n+1|+h.c]\}\nonumber\\
\end{eqnarray}
which leads to the logarithmic negativity of the above density
matrix
\begin{eqnarray}
E(\rho)&=&\log_2{[1+\sum_{n=0}^\infty(|\xi^-_n|-\xi^-_n)]},\nonumber\\
\xi_n^-&=&\frac{1}{2}\{p_{n-1}|\beta_n|^2+p_{n+1}|\alpha_{n+1}|^2\nonumber\\
&&-[(p_{n-1}|\beta_n|^2-p_{n+1}|\alpha_{n+1}|^2)^2\nonumber\\
&&+4|p_n\alpha_n\beta^*_{n+1}|^2]^{1/2}\}.
\end{eqnarray}
The logarithmic negativity is plotted as a function of time t in
Fig.1 that when the field is initially prepared in the vacuum state
$|0\rangle$ or in the Fock state $|1\rangle$.  For the sake of
simplicity, we set $g=1$ in the numerical calculation. We can see
clearly from Fig.1, the atom-field entanglement can be increased by
applying the classical driving field.

\begin{figure}
\centering {\includegraphics{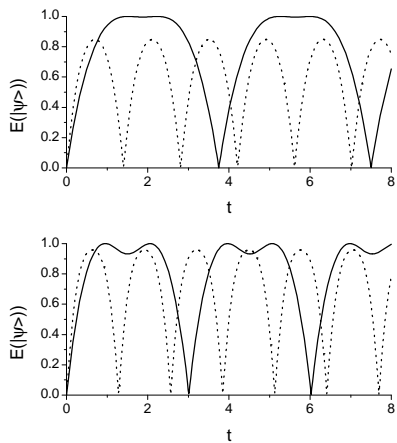}}
 \caption{
  The logarithmic negativity $E(|\psi\rangle)$ is plotted as a function of t
  with $\omega=5, \omega_0=1$ for $\omega_c=\lambda=0$(dotted line) and $\omega_c=\lambda=2$(solid line).
  Upper panel: The field is initially prepared in the vacuum state. Lower
  panel: The field is initially prepared in the Fock state
  $|1\rangle$.
  }
\end{figure}

In Fig.2, the logarithmic negativity $E(\rho)$ is plotted as a
function of t with $\omega=5, \omega_0=1$ for
$\omega_c=\lambda=0$(dotted line) and $\omega_c=\lambda=2$(solid
line). The mean photon number $\overline{m}$ for the upper panel and
the lower panel are 0.1 and 0.3, respectively. It is not difficult
to find that the entanglement of the two-atom system decreases with
the mean photon number $\overline{m}$. In other words, the amount of
entanglement decreases with the increase of the temperature of the
thermal field. However, we can increase the amount of entanglement
by controlling the classical field.

\begin{figure}
\includegraphics{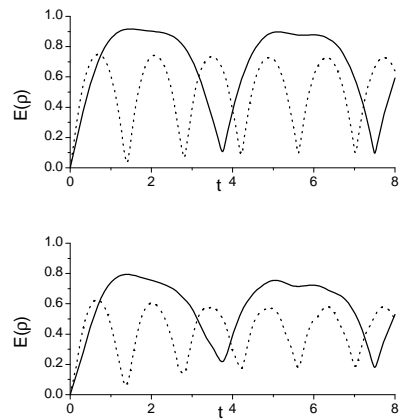}
 \caption{
  The logarithmic negativity $E(\rho)$ is plotted as a function of t
  with $\omega=5, \omega_0=1$ for $\omega_c=\lambda=0$(dotted line) and
$\omega_c=\lambda=2$(solid line). Upper panel: The mean photon
number of the thermal field is 0.1, i.e., $\overline{m}=0.1$. Lower
panel: The mean photon number of the thermal field is 0.3, i.e.,
$\overline{m}=0.3$.}
\end{figure}

\section{Controlling entanglement sudden death }
In the previous section, we have investigated the entanglement
dynamics of a two-level atom interacting with a cavity field.
Recently, Bellomo \emph{et al}. \cite{Bellomo2007, Bellomo2008}
investigated the influence of the Markovian and Non-Markovian
effects on the dynamics of two-qubit entanglement. In this section,
we consider a quantum system consisting of two noninteracting atoms
each locally interacting with its own vacuum field. The two atoms,
which are initially prepared in EWL states, are driven by two
independent classical fields additionally. Our purpose here is to
investigate the influence of the classical driving field and the
purity of the initial states on the entanglement dynamics of the
system.

\subsection{Reduced density matrix of the two-atom system}
Here, we study the entanglement dynamics of two independent atoms
each locally interacting with a vacuum field by using the procedure
of Ref.\cite{Bellomo2008}. In addition, each atom is driven by a
classical field. In the following, we find the reduced density
matrix of the two-atom system. In the basis $|\pm\rangle$, the
initial state of the atom interacting with the vacuum field can be
cast in the following form
\begin{eqnarray}
\left(
\begin{array}{cc}
  \rho_{++}(0) & \rho_{+-}(0) \\
  \rho_{-+}(0) & \rho_{--}(0)
  \end{array}
  \right)\otimes|0\rangle\langle0|,
\end{eqnarray}
where $|+\rangle$ and $|-\rangle$ are defined by Eq.(8). Using the
state vector in Eq.(8) and tracing over the freedom of the field, we
obtain the single-atom density matrix evolution
\begin{eqnarray}
\left(
\begin{array}{cc}
  \rho_{++}(t) & \rho_{+-}(t) \\
  \rho_{-+}(t) & \rho_{--}(t)
  \end{array}
  \right),
\end{eqnarray}
where
\begin{eqnarray}
\rho_{++}(t)&=&|\alpha_0(t)|^2\rho_{--}(0),\nonumber\\
\rho_{--}(t)&=&[1-|\alpha_0(t)|^2]\rho_{++}(0),\nonumber\\
\rho_{+-}(t)&=&\alpha_0(t)\rho_{+-}(0)=\rho^*_{-+}(t).
\end{eqnarray}
By making use of the above equations, we can find the dynamics of
the two atoms via a purely algebraic way. Note that this procedure
is applicable to arbitrary initial states of the whole system. In
the basis $|1\rangle=|++\rangle, |2\rangle=|+-\rangle,
|3\rangle=|-+\rangle, |4\rangle=|--\rangle,$ and using Eq.(8) and
Eq.(15), the density matrix for the two-atom system is calculated as
follows
\begin{eqnarray}
\rho_{11}(t)&=&|\alpha_0(t)|^4\rho_{11}(0),\nonumber\\
\rho_{22}(t)&=&|\alpha_0(t)|^2[1-|\alpha_0(t)|^2]\rho_{11}(0)+|\alpha_0(t)|^2\rho_{22}(0),\nonumber\\
\rho_{33}(t)&=&|\alpha_0(t)|^2[1-|\alpha_0(t)|^2]\rho_{11}(0)+|\alpha_0(t)|^2\rho_{33}(0),\nonumber\\
\rho_{44}(t)&=&[1-|\alpha_0(t)|^2]^2\rho_{11}(0)+\rho_{44}(0)\nonumber\\
&&+[1-|\alpha_0(t)|^2][ \rho_{22}(0)+\rho_{33}(0)],\nonumber\\
\rho_{12}(t)&=&\alpha_0(t)|\alpha_0(t)|^2\rho_{12}(0),\nonumber\\
\rho_{13}(t)&=&\alpha_0(t)|\alpha_0(t)|^2\rho_{13}(0),\nonumber\\
\rho_{14}(t)&=&\alpha_0(t)^2\rho_{14}(0),\nonumber\\
\rho_{23}(t)&=&|\alpha_0(t)|^2\rho_{23}(0),\nonumber\\
\rho_{24}(t)&=&\alpha_0(t)[1-|\alpha_0(t)|^2]\rho_{13}(0)+\alpha_0(t)\rho_{24}(0),\nonumber\\
\rho_{34}(t)&=&\alpha_0(t)[1-|\alpha_0(t)|^2]\rho_{12}(0)+\alpha_0(t)\rho_{34}(0),
\end{eqnarray}
with $\rho_{ij}(t)=\rho^*_{ji}(t)$. We would like to point out that
the above procedure allows us to obtain the reduced density matrix
of the two-atom system for any initial state.

\subsection{Extended Werner-like states and logarithmic negativity}
We assume the initial states of the two-atom system are the extended
Werner-like states defined by
\begin{eqnarray}
\rho_{\Phi}(0)&=&r|\Phi\rangle\langle
\Phi|+\frac{1-r}{4}I,\nonumber\\
\rho_{\Psi}(0)&=&r|\Psi\rangle\langle
\Psi|+\frac{1-r}{4}I,\nonumber\\
|\Phi\rangle&=&\mu|-+\rangle+\nu |+-\rangle,\nonumber\\
|\Psi\rangle&=&\mu |--\rangle+\nu|++\rangle,
\end{eqnarray}
where r is a real number which indicates the purity of initial
states, $I$ is a $4\times 4$ identity matrix, $\mu$ and $\nu$ are
complex numbers with $|\mu|^2+|\nu|^2=1$. It is noted that the EWL
states belong to the class of the `X' states. Explicitly, if the
density matrix of a quantum state is of the form
\begin{eqnarray}
\left(
\begin{array}{cccc}
  \rho_{11}        &    0          & 0           & \rho_{14}\\
    0              & \rho_{22}     &\rho_{23}    &   0\\
     0             & \rho^*_{23}    &\rho_{23}    &   0\\
      \rho^*_{14}  & 0              &0           &   \rho_{44}\\
  \end{array}
  \right),
\end{eqnarray}
then it belongs to the class of the X states.

The EWL states have the following advantages. First, we can easily
find that the density matrix at arbitrary time t is still X
structure under the single atom evolution determined by the
Hamiltonian $H_2$ (in the basis $|1\rangle=|++\rangle,
|2\rangle=|+-\rangle, |3\rangle=|-+\rangle, |4\rangle=|--\rangle$)
if the initial state is X structure, such as the Bell states and the
EWL states. Second, the EWL states allow us to clearly show the
influence of the purity and the amount of entanglement of the
initial states on the entanglement dynamics simultaneously. The
purity of the EWL states are dependent on the parameter r and the
amount of the entanglement of the EWL states are related to $r$ and
$\mu$. If $r=1$ the EWL states reduce to the Bell-like states
$|\Phi\rangle$ and $|\Psi\rangle$. In the case of
$r=1,\mu=\nu=1/\sqrt{2}$ the EWL states become the Bell states while
in the case of $r=0$ they are the maximally  mixed states.

Note that we assume that the initial states of the system are the
EWL states. The X structure of the EWL states is maintained during
the evolution of the two-atom system. The explicit analytical
expression of the logarithmic negativity for the two-atom system can
be obtained as
\begin{eqnarray}
E_X&=&\log_2(1+2N),\nonumber\\
N&=&\max\{0,\frac{1}{2}[\sqrt{(\rho_{22}-\rho_{33})^2+4|\rho_{14}|^2}-\rho_{22}-\rho_{33}]\}\nonumber\\
&&+\max\{0,\frac{1}{2}[\sqrt{(\rho_{11}-\rho_{44})^2+4|\rho_{23}|^2}\nonumber\\
&&-\rho_{11}-\rho_{44}]\}.
\end{eqnarray}
 We plot the
logarithmic negativity as a function of time for different values of
the purity of the initial states r and the strength of the classical
field $\lambda$. In what follows, we denote the entanglement of the
two-atom system which is initially prepared in $\rho_{\Phi}(0)$
($\rho_{\Psi}(0))$ by $E_{\Phi}$ ($E_{\Psi}$).

\begin{figure}
\includegraphics{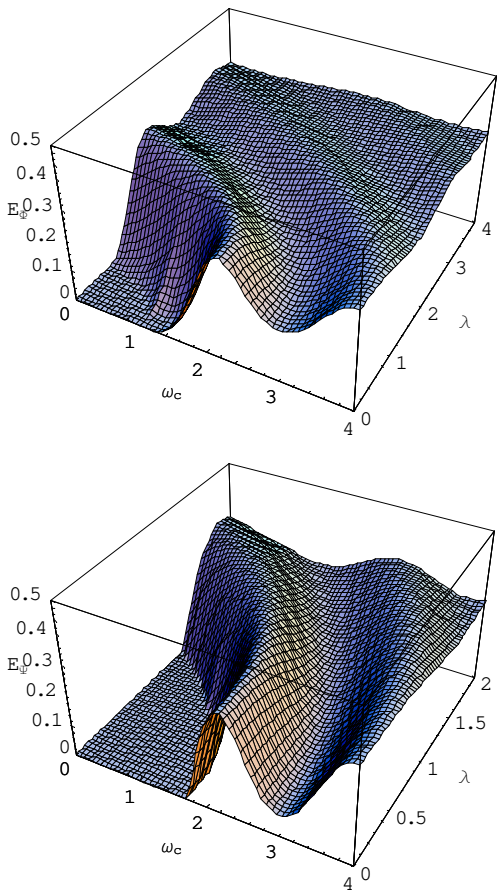}
 \caption{
 Upper panel:  The logarithmic negativity $E_{\Phi}$ is plotted as a function of
 $\lambda$ and $\omega_c$
  with $\omega=\omega_0=1, |\mu|^2=1/6, |\nu|^2=5/6, r=2/3, t=2$.
  Lower panel: The logarithmic negativity $E_{\Psi}$ is plotted as a function of
 $\lambda$ and $\omega_c$
  with $\omega=\omega_0=1, |\mu|^2=1/6, |\nu|^2=5/6, r=2/3, t=2$.}
\end{figure}

The logarithmic negativity $E_{\Phi}$ and $E_{\Psi}$ are plotted as
a function of $\lambda$ and $\omega_c$ with $\omega=\omega_0=1,
|\mu|^2=1/6, |\nu|^2=5/6, r=2/3, t=2$ in the upper panel and the
lower panel of Fig.3, respectively. From Fig.3, one can clearly see
that if the parameters $\omega_c$ and $\lambda$ are small, then the
logarithmic negativity $E_\Phi$ and $E_\Psi$ are zero at time $t=2$.
The situation is different if the parameters of the classical
driving fields increase. When the values of $\omega_c$ and $\lambda$
are lager enough the two atoms become entangled. The behavior of
$E_\Phi$ and $E_\Psi$ are different when the classical fields are
applied. For example, the area for $E_\Phi=0$ is small than that of
$E_\Psi=0$ which means that the entanglement of the two-atom system
is more sensitive with the classical driving fields if they are
initially prepared in $\rho_\Phi$. This feature can be see more
clearly in Fig.4 and Fig.5. It also interesting to note that the
entanglement of the two-atom system can be significantly increased
by controlling the classical driving fields both for the initial
states $\rho_\Phi$ and $\rho_\Psi$.

\begin{figure}
\includegraphics{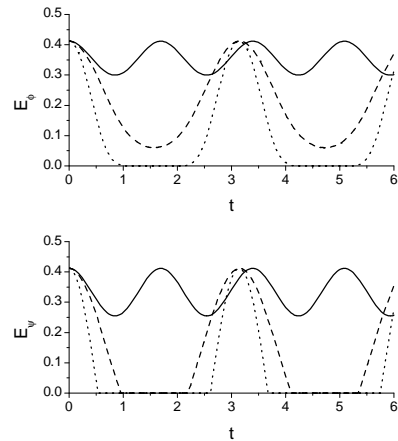}
 \caption{Upper panel: The logarithmic negativity $E_{\Phi}$ is plotted as a function of
 t with $|\mu|^2=1/6, |\nu|^2=5/6, r=2/3, \omega=\omega_0=1$ for $\omega_c=\lambda=0$
(dotted line), $\omega_c=0.5, \lambda=1$ (dashed line), and
$\omega_c=0.5, \lambda=2$ (solid line).
  Lower panel: The logarithmic negativity $E_{\Psi}$ is plotted as a function of
 t with $|\mu|^2=1/6, |\nu|^2=5/6, r=2/3, \omega=\omega_0=1$ for $\omega_c=\lambda=0$
(dotted line), $\omega_c=0.5, \lambda=1$ (dashed line), and
$\omega_c=0.5, \lambda=2$ (solid line).}
\end{figure}

In Fig.4, we plot the logarithmic negativity of the two-atom system
as a function of time t for several values of $\omega_c$ and
$\lambda$. The upper panel of Fig.4 shows that the ESD occurs
without the classical field (see the dotted line in this panel).
After finite dark periods, the entanglement $E_{\Phi}$ revivals
completely. The entanglement can be increased by applying the
classical field(see the dashed and solid lines of the panel). It is
interesting to point out that the ESD disappears when the classical
field is strong enough. Comparing the upper panel with the lower
panel of Fig.4, we find that the behavior of entanglement $E_{\Phi}$
and $E_{\Psi}$ are different since the ESD appears in $E_{\Psi}$
with $\omega_c=0.5, \lambda=1$ while there is no ESD for $E_{\Phi}$
with the same values of $\omega_c$ and $\lambda$. Besides, the
entanglement dark periods of $E_{\Psi}$ is longer than that of
$E_{\Phi}$ when all of the other correspongding parameters are the
same. In other words, it takes more time for $E_{\Psi}$ to revive
the initial entanglement. The parameters $\mu$ and $\nu$ in Fig.4
are chosen to be $\sqrt{1/6}$ and $\sqrt{5/6}$, respectively. It is
worth noting that different choices of $\mu$ and $\nu$ do not give
entanglement dynamics qualitatively different from the case treated
here.

Another aspect of interest is how the entanglement dynamics is
influenced by the purity of the initial states. In order to show it
intuitively, we plot the logarithmic negativity of the two-atom
system as a function of t for different values of r in Fig.5, fixing
the parameters $\mu$ and $\nu$. From Fig.5, one can see that the
entanglement dark periods increase with the decrease of the purity
of the initial states. The classical driving field again can
increase the amount of the entanglement of the system.

\begin{figure}
\includegraphics{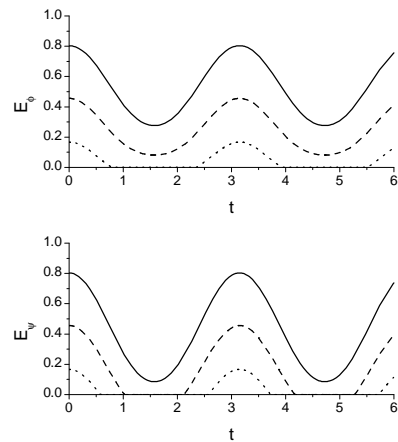}
 \caption{Upper panel: The logarithmic negativity $E_{\Phi}$ is plotted as a function of
 t with $|\mu|^2=1/6, |\nu|^2=5/6, \omega=\omega_0=1, \omega_c=0.5, \lambda=1 $ for $r=0.5$
(dotted line), $r=0.7$ (dashed line), and $r=1$ (solid line).
  Lower panel: The logarithmic negativity $E_{\Psi}$ is plotted as a function of
 t with $|\mu|^2=1/6, |\nu|^2=5/6, \omega=\omega_0=1, \omega_c=0.5, \lambda=1 $ for $r=0.5$
(dotted line), $r=0.7$ (dashed line), and $r=1$ (solid line).}
\end{figure}

\section{CONCLUSIONS}
In the present paper, we have considered a quantum system consisting
of one two-level atom interacting with a single mode field. The atom
is driven by a classical field additionally. The entanglement
dynamics is described by the logarithmic negativity. We first
investigated the entanglement dynamics of the atom-field system. The
field is initially prepared in the Fock state or the thermal state.
We find that the atom-field entanglement can be increased by
applying the classical driving field.

Then, we consider a quantum system consisting of two noninteracting
atoms each locally interacting with its own vacuum field. The two
atoms, which are driven by two independent classical fields, are
initially prepared in EWL states, i.e., $\rho_{\Phi}$ and
$\rho_{\Psi}$. We find that classical driving fields can increase
the amount of entanglement of the two-atom system. The behavior of
entanglement $E_\Phi$ and $E_\Psi$ are different when the classical
fields are applied. The entanglement of the two-atom system is more
sensitive with the classical driving fields if they are initially
prepared in $\rho_\Phi$. It is worth noting that the time of
entanglement sudden death can be controlled by the classical driving
fields. Specially, the entanglement sudden death phenomenon will
disappear if the classical driving fields are strong enough. It is
interesting to verify the scheme of our paper in cavity QED
experimentally.

\section*{ ACKNOWLEDGEMENTS}
 This project was supported by the National Natural
Science Foundation of China (Grant no.10774131) and the National Key
Project for Fundamental Research of China (Grant no. 2006CB921403).

\bibliographystyle{apsrev}

\begin{thebibliography}{}

\bibitem{Cirac2000}
 J. I. Cirac, P. Zoller,  Nature {\bf 404}, 579-581(2000).

\bibitem{Nielsen2000}
 M.A. Nielsen, I. L. Chuang, {\it Quantum Computation and Quantum
Information} (Cambridge University Press, Cambridge, 2000).

\bibitem{Bennett1993} C. H. Bennett, G. Brassard, C. Crepeau,
R. Jozsa, A. Peres, W. K. Wootters, Phys. Rev. Lett. {\bf 70},
1895-1899 (1993).

\bibitem{vedral2000} V. Vedral, M. B. Plenio, Phys. Rev. A {\bf
57}, 1619-1633 (2000).

\bibitem{Raimond2001} J. M. Raimond, M. Brune, S. Haroche, Rev. Mod.
Phys {\bf 73},  565-582 (2001).

\bibitem{Bose2001} S. Bose, I. Fuentes-Guridi, P. L. Knight,
 V. Vedral, Phys. Rev. Lett. {\bf 87},  050401 (2001).

\bibitem{Kim2002} M. S. Kim, Jinhyoung  Lee, D. Ahn,
 P. L. Knight, Phys. Rev. A {\bf 65},  040101(R) (2002).

\bibitem{Paternostro2007} M. Paternostro, D. Vitali, S. Gigan, M. S. Kim, C. Brukner,
J. Eisert, M. Aspelmeyer, Phys. Rev. Lett. {\bf 99},  250401 (2007).

\bibitem{Vitali2007}
D. Vitali, S. Gigan, A. Ferreira, H. R. Bohm, P. Tombesi, A.
Guerreiro, V. Vedral, A. Zeilinger, M. Aspelmeyer, Phys. Rev. Lett.
{\bf 98},  030405 (2007).

\bibitem{Li2005} S. B. Li and  J. B. Xu, Phys. Rev. A {\bf
72},  022332 (2005).

\bibitem{Marchiolli2006}  M. A. Marchiolli, J. Mod. Opt.
{\bf 53},  2733-2751 (2006).

\bibitem{Ke2005} S. S.  Ke and  G. X. Li, J. Mod. Opt.
{\bf 52},   2743-2758(2005).

\bibitem{Braun2002} Braun D., Phys. Rev. Lett. {\bf 89},  277901 (2002).

\bibitem{Xu2005}J. B. Xu and S. B. Li, New. J. Phys., {\bf 7}, 72 (2005).


\bibitem{Solano2003}
 E. Solano, G. S. Agarwal, and H. Walther,
 Phys. Rev. Lett. {\bf 90 },
027903  (2003).


\bibitem{Yu2004}
Y. Yu and J. H. Eberly,  Phys. Rev. Lett. {\bf 93}, 140404  (2004).

\bibitem{Jakobczyk2004}
L. Jak\'{o}bczyk and A. Jammr\'{o}z,  Phys. Lett. A {\bf  333}, 35
(2004).

\bibitem{Ficek2006}
Z. Ficek and R. Tana\'{s},  Rev. Rev. A {\bf 74}, 024304 (2006).

\bibitem{Yonac}
M. Yonac, T. Yu, and J. H. Eberly, arXiv: quant-ph/0602206.

\bibitem{Yu2006}
Y. Yu and J. H. Eberly,  Opt. Commun {\bf 264}, 393 (2006).

\bibitem{Ikram2008} M. Ikram, Fu-Li Li, and M. S. Zubairy,
Phys. Rev. A {\bf 75}, 062336 (2007).

\bibitem{Bellomo2007} B. Bellomo, R. Lo France, and G. Compagno,
Phys. Rev. Lett. {\bf 99}, 160502 (2007).

\bibitem{Bellomo2008} B. Bellomo, R. Lo France, and G. Compagno,
Phys. Rev. A {\bf 77}, 032342 (2008).

\bibitem{Almeida2007} M. P. Almeida, F. de Melo, M. Hor-Meyll, A. Salles, S. P. Walborn,
P. H. Souto Ribeiro, L. Davidovich, Science(London) {\bf 316}, 579
(2007).

\bibitem{Vidal2002} G. Vidal and R. F. Werner, Phys. Rev. A {\bf
65} 032314 (2002).

\bibitem{Plenio2005} M. B. Plenio, Phys. Rev. Lett. {\bf 95}, 090503
(2005).


\bibitem{Liu2006} Yu-Xi Liu, C. P. Sun, and Franco Nori, Phys. Rev.
A {\bf 74}, 052321 (2006).

\bibitem{Vedral1997}
  V. Vedral, M. B. Plenio,
M. A. Rippin, and P. L. Knight, Phys. Rev. Lett. {\bf 78}, 2275
(1997).

\end{thebibliography}

\end{document}